\documentclass[11pt]{article}
\usepackage[paper=a4paper,dvips,top=2.0cm,left=2.2cm,right=2.2cm,bottom=2.6cm,footskip=2.2cm,voffset=0.0cm, marginparwidth=18mm]{geometry}
\usepackage{amsfonts,amsmath,amsthm,amssymb}
\numberwithin{equation}{section} 
\usepackage[svgnames]{xcolor}
\usepackage{fix-cm}
\usepackage{sectsty}
\usepackage{fancyhdr}
\usepackage{lastpage}
\usepackage{graphicx}
\usepackage{url}
\usepackage{enumerate}
\usepackage{paralist}
\usepackage{multicol}
\usepackage{color}  
\usepackage{float}  
\usepackage{epsfig}
\usepackage{subfigure}
\usepackage{hyperref}   
\hypersetup{colorlinks=true,linkcolor=beamer@PRD, citecolor=beamer@PRD}
\usepackage{authblk}  
\usepackage{cite}  
\usepackage{todonotes}
\usepackage{ulem} % Enables sout command
\newcommand*\doperator{\mathop{}\!\mathrm{\,d}}
\providecommand*{\deriv}[3][]{\frac{\doperator^{#1}#2}{\doperator #3^{#1}}}
\providecommand*{\pderiv}[3][]{\frac{\partial^{#1}#2}{\partial #3^{#1}}}
\renewcommand\eqref[1]{\textcolor{beamer@PRD}{(}\ref{#1}\textcolor{beamer@PRD}{)}}
\definecolor{beamer@PRD}{RGB}{46,48,146}
\begin{document}
%%%%%%%%%%%%%%%%%%%%%%%%%%%%%%%%%%%%%%%%%%%%%%%%%%%%%%%%%%%%%%%%%%%%%%%%%%%%%%
%  Title
%%%%%%%%%%%%%%%%%%%%%%%%%%%%%%%%%%%%%%%%%%%%%%%%%%%%%%%%%%%%%%%%%%%%%%%%%%%%%%
\title{\textbf{Dynamic noncommutative BTZ black holes}}
\author{\textbf{Ankur$^{1,2}$ and Sanjib Dey$^{1,\ast}$} \\ \small{${}^1$Department of Physical Sciences, Indian Institute of Science Education and Research Mohali, \\ Sector 81, SAS Nagar, Manauli 140306, India\\ ${}^2$Dipartimento di Fisica, Universit\`a degli Studi di Trieste \\ Strada Costiera 11, 34151 Miramare--Trieste, Italy \\ $^\ast$E-mail: sanjibdey4@gmail.com}}
\date{}
\maketitle
%%%%%%%%%%%%%%%%%%%%%%%%%%%%%%%%%%%%%%%%%%%%%%%%%%%%%%%%%%%%%%%%%%%%%%%%%%%%%%
%  Abstract
%%%%%%%%%%%%%%%%%%%%%%%%%%%%%%%%%%%%%%%%%%%%%%%%%%%%%%%%%%%%%%%%%%%%%%%%%%%%%%
%\thispagestyle{fancy}
\begin{abstract}
We have studied the charged BTZ black holes in noncommutative spaces arising from two independent approaches. First, by using the Seiberg-Witten map followed by a dynamic choice of gauge in the Chern-Simons gauge theory. Second, by inducing the fuzziness in the mass and charge by a Lorentzian distribution function with the width being the same as the minimal length of the associated noncommutativity. In the first approach, we have found the existence of non-static and non-stationary BTZ black holes in noncommutative spaces for the first time in the literature, while the second approach facilitates us to introduce a proper bound on the noncommutative parameter so that the corresponding black hole becomes stable and physical. We have used a contemporary tunneling formalism to study the thermodynamics of the black holes arising from both of the approaches and analyze their behavior within the context.
\end{abstract}	 
%%%%%%%%%%%%%%%%%%%%%%%%%%%%%%%%%%%%%%%%%%%%%%%%%%%%%%%%%%%%%%%%%%%%%%%%%%%%%%
%  Introduction
%%%%%%%%%%%%%%%%%%%%%%%%%%%%%%%%%%%%%%%%%%%%%%%%%%%%%%%%%%%%%%%%%%%%%%%%%%%%%%
\section{Introduction} \label{sec1}
\addtolength{\footskip}{-0.2cm} %% Add extra space in the first page of the article
%%%%%%%%%%%%%%%%%%%%%%%%%%%%%%%%
It is argued that at a very high energy scale, the fundamental concept of smooth space-time geometry no longer holds due to the fact that the quantum fluctuation in such a scale contributes significantly, which cannot be dealt with a perturbation around the classical geometry. Though we have not yet completely conceived a fully operating quantum geometry, among various propositions, noncommutative geometry is believed to apprehend some of the requisite characteristics. Since then the noncommutative geometry has become a central area of interest, which; essentially replaces the pointlike concepts with smeared objects \cite{Connes,Garay,Seiberg_Witten}. Noncommutativity plays a crucial role in many approaches of quantum gravity \cite{Douglas_Hull,Amelino,Kowalski} as well as in several other branches of modern physics \cite{Bellissard,Gopakumar,Calmet,Dey_Fring_PRD,Dey_Fring_Gouba_Castro_PRD,anacleto, Dey_Fring_2014,Dey_cat, Dey_Hussin,bourne,Dey_Hussin_PRA, dey2017, Dey_Fring_Hussin_2017, Dey_atm_laser}. A particular interest is dispersed among different contexts of general relativity; such as wormholes \cite{sharif} and black-holes \cite{nicolini2005,nicolini2006noncommutative,Rizzo,ansoldi2007non, giri,Nozari2008,kim2008,chang2009,spallucci2009,larranaga2010three,ding2011, mann2011,liang2012thermodynamics, Rahaman2013btz,hendi2015,kawamoto2018charged,anacleto2018quantum}.

The first solution of noncommutative Schwarzschild black hole in 4-dimensions was found in \cite{nicolini2005,nicolini2006noncommutative}. Later, the model was prolonged to incorporate extra spatial dimensions \cite{Rizzo,spallucci2009} as well as the electric charge \cite{ansoldi2007non}. Different scenarios of noncommutative black holes in diverse contexts have been explored over the years, such as the pair creation in a positive cosmological constant background \cite{mann2011}, asymptotic quasinormal modes \cite{giri}, Parikh-Wilczek tunneling \cite{Nozari2008}, etc. The 2-dimensional \cite{ding2011} and 3-dimensional \cite{kim2008,chang2009, liang2012thermodynamics,Rahaman2013btz,hendi2015,kawamoto2018charged,anacleto2018quantum} studies of noncommutative black holes have also attracted a lot of interest in recent days.

Among various ways of implementing the noncommutativity in black holes, the most notable approaches include the application of the Seiberg-Witten (SW) map \cite{Seiberg_Witten} on the Chern-Simons (CS) gauge theory \cite{chang2009,kawamoto2018charged} and the inclusion of noncommutative smearing by replacing point-like sources with a Gaussian \cite{nicolini2006noncommutative,larranaga2010three, Rahaman2013btz} or a Lorentzian distribution \cite{Rizzo,Nozari2008,liang2012thermodynamics}. In this article, we shall study charged BTZ black holes in noncommutative spaces by using both of the above approaches. In particular, we shall provide non-static and non-stationary BTZ black hole solutions by using the first approach. It should be noted that the existence of non-static \cite{Ghosh2012, Ahmed} and non-stationary \cite{hayward1999,alcubierre2001,wu2002} black holes were found earlier, however, all of them arise from the standard commutative space-time background. Here, we shall explore them in the noncommutative space-time, which; to our knowledge were not studied earlier. By following the second approach, we shall construct a metric of a charged noncommutative BTZ black hole from a fuzzy Lorentzian distribution. The main motivation of studying both of the approaches sides by side is that, while the first approach gives rise to some interesting unfamiliar black holes, the other approach reinforces us to provide an explicit bound on the noncommutative parameter, which itself is a fascinating area of research in noncommutative theories. Moreover, by analyzing the thermodynamic behavior in the tunneling formalism, we realize that the bound on noncommutativity is a consequence of the addition of charge to the black hole. This is why we believe that unveiling exotic characteristics of charged black holes in noncommutative space can be impactful and rewarding.

The paper is organized as follows. In Sec.\,\ref{sec2}, we introduce the basic notations and recall the detailed technicalities of two of the mechanisms of our interest for constructing noncommutative BTZ black holes. In the rest of the sections, we discuss our main results and findings. In Sec.\,\ref{sec3}, we explore the non-static and non-stationary BTZ black hole solutions resulting from a unique dynamic choice of gauge in CS theory. Sec.\,\ref{sec4} is composed of a charged noncommutative BTZ black hole solution with a Lorentzian distribution function as well as a pragmatic discussion on the bound of the noncommutative parameter. In Sec.\,\ref{sec5}, we analyze the thermodynamics of the black holes which were constructed in earlier sections. Finally, our conclusions are stated in Sec.\,\ref{sec6}.
%%%%%%%%%%%%%%%%%%%%%%%%%%%%%%%%%%%%%%%%%%%%%%%%%%%%%%%%%%%%%%%%%%%%%%%%%%%%%%
% Section 2
%%%%%%%%%%%%%%%%%%%%%%%%%%%%%%%%%%%%%%%%%%%%%%%%%%%%%%%%%%%%%%%%%%%%%%%%%%%%%%
\section{BTZ black holes in noncommutative space}\label{sec2}
\lhead{Dynamic noncommutative BTZ black holes}
\chead{}
\rhead{}
\addtolength{\voffset}{0.8cm} %% Adds extra space in the header of the 1st page of the article 
\addtolength{\footskip}{-0.7cm} %% Adds extra space in the first page of the article 
%%%%%%%%%%%%%%%%%%%%%%%%%%%%%%%%%%
In this section, we shall discuss the construction method of BTZ black holes in noncommutative space arising from two independent approaches. In the first approach, we shall utilize the SW map \cite{Seiberg_Witten} to obtain the noncommutative version of the CS gauge theory, which when applied to reconstruct the veilbeins and spin connections, one obtains the metric of the required black hole. In the second approach, we shall incorporate the noncommutativity in a more natural way by replacing the point-like concept of a particle with the idea of smearing of the object over space. In other words, instead of representing the mass and charge of a particle by a Dirac delta function, one can express them by a Gaussian or a Lorentzian function whose width is in the order of the minimal length of the underlying noncommutative space.

\subsection{Using CS gauge theory and SW map}\label{sec21}
In the first-order formulation of gravity in $(2+1)$ dimensions, the Einstein-Hilbert action $\mathcal{I}_{\text{EH}}$ with nonzero vacuum energy becomes equivalent \cite{Achucarro_Townsend,Witten} to the action functional of the CS theory
\begin{equation}
\mathcal{S}_{\text{CS}}=\mathcal{I}_{\text{CS}}[\mathcal{A^{(+)}}]-\mathcal{I}_{\text{CS}}[\mathcal{A^{(-)}}]\equiv\mathcal{I}_\text{EH},
\end{equation}
with two $SU(1,1)\simeq SO(1,2)$ connection $1$-forms
\begin{equation}\label{CSgauge}
\mathcal{A}^{(\pm)a}=\omega^a\pm\frac{e^a}{\ell},
\end{equation}
where the CS term, $\mathcal{I}_{\text{CS}}=k/4\pi\int \text{tr}[\mathcal{A}\doperator\mathcal{A}+2\mathcal{A}\mathcal{A}\mathcal{A}/3]$, is given in terms of the CS 3-form, Here, $k=-\ell/4G$ represents the CS level, and $e^a$ and $\omega^a$ denote the vielbeins and spin connections, respectively, for $a=0,1,2$. Here, $G$ is the universal gravitational constant and $\ell$ is the length scale associated with the cosmological constant $\Lambda=-1/\ell^{2}$. The metric of a charged rotating BTZ black hole is given by \cite{BTZ}
\begin{equation}\label{CRBTZ}
ds^{2}=-f(r)\doperator t^{2}+\frac{\doperator r^{2}}{g(r)}+r^{2}\doperator\varphi^{2}-\frac{2 \gamma}{\ell} \doperator t\doperator \varphi,
\end{equation}
with
\begin{equation}
g(r)=f(r)+\frac{\gamma^{2}}{\ell^{2} r^{2}},~\quad f(r)=\frac{1}{\ell^{2}}\left(r^{2}-\zeta-\epsilon \ln r\right),~\zeta=8 G M \ell^{2},~\epsilon =8 \pi G \ell^{2} Q^{2},~\gamma=4 G J \ell,
\end{equation}
where, $M,Q,J$ denote the total mass, charge and angular momentum of the black hole, respectively. One can choose a set of non-unique vielbeins
\begin{equation}\label{vielbein}
e^0=\sqrt{f(r)}\doperator t+\frac{\gamma}{\ell\sqrt{f(r)}}\doperator\varphi, \quad e^1=\frac{\doperator r}{\sqrt{g(r)}}, \quad e^2=r\sqrt{\frac{g(r)}{f(r)}}\doperator\varphi,
\end{equation}
to rewrite the metric \eqref{CRBTZ} in terms of the vielbeins as $ds^2=-(e^0)^2+(e^1)^2+(e^2)^2$. The corresponding spin connections are given by
\begin{equation}\label{spinc}
\omega^0=-\frac{\gamma f'(r)}{2\ell r\sqrt{f(r)}}\doperator t-\sqrt{f(r)}\doperator\varphi, \quad \omega^1=\frac{\gamma f'(r)}{2\ell rf(r)\sqrt{g(r)}}\doperator r, \quad \omega^2=-\frac{f'(r)}{2}\sqrt{\frac{g(r)}{f(r)}}\doperator t.
\end{equation}
It is worth mentioning that throughout the manuscript, we denote $(\cdot)'$ as the first-order derivative with respect to $r$. A noncommutative counterpart of the charged rotating BTZ black hole can be obtained by using the SW map \cite{Seiberg_Witten} for gauge fields, where the noncommutativity between the radial and angular variables is imposed through the commutation relation \cite{chang2009}
\begin{equation}\label{SW_Map}
[\mathcal{R},\varphi]=2i\theta, \quad \mathcal{R}=r^2,
\end{equation}
where, $\theta^{R\varphi}=-\theta^{\varphi R}=2\theta$ for a constant $\theta$, with all other components being zero. The SW map imposes the compatibility condition for gauge transformation between $\mathcal{A}$ and $\tilde{\mathcal{A}}$
\begin{equation}
\tilde{\mathcal{A}}(\mathcal{A})+\tilde{\delta}_{\tilde{\xi}}\tilde{\mathcal{A}}(\mathcal{A})=\tilde{\mathcal{A}}(\mathcal{A}+\delta_\xi\mathcal{A}),
\end{equation}
which when solved using the perturbation theory for infinitesimal $\xi$ and $\tilde{\xi}$, we obtain
\begin{eqnarray}
&&\tilde{\mathcal{A}}_{\mu}(\mathcal{A})=\mathcal{A}_{\mu}-\frac{i}{4} \theta^{v \rho}\left\{\mathcal{A}_{v}, \partial_{\rho} \mathcal{A}_{\mu}+\mathcal{F}_{\rho \mu}\right\}+\mathcal{O}\left(\theta^{2}\right), \\
&&\tilde{\xi}(\xi,\mathcal{A})=\xi+\frac{i}{4}\theta^{\mu\nu}\{\partial_\mu\xi,\mathcal{A}_\nu\}+\mathcal{O}(\theta^{2}).
\end{eqnarray}
Here, $\{\cdot,\cdot\}$ and $\mathcal{F}_{\mu\nu}$ represent the anticommutator with respect to the conventional matrix product and the field strength for the gauge fields $\mathcal{A}_\mu$, respectively. For the noncommutativity of the form \eqref{SW_Map}, the correction terms to the gauge fields are given by\cite{kawamoto2018charged}
 \begin{equation}\label{2}\begin{aligned}
\mathbb{A}_{\mu}(\mathcal{A})=&-\frac{i\theta}{2}\left[\frac{1}{2} \eta_{ab} \mathcal{A}_{R}^{a}\left(\partial_{\varphi} \mathcal{A}_{\mu}^{b}+\mathcal{F}_{\varphi \mu}^{b}\right) \mathbb{I}-\frac{1}{2} \eta_{ab} \mathcal{A}_{\varphi}^{a}\left(\partial_{R} \mathcal{A}_{\mu}^{b}+\mathcal{F}_{R \mu}^{b}\right) \mathbb{I}\right.\\
&+i\left(\mathcal{A}_{R}^{a} \tau_{a}+\mathcal{B}_{R} \tau_{3}\right)\left(\partial_{\varphi} \mathcal{B}_{\mu}+\mathcal{F}_{\varphi \mu}^{(\mathcal{B})}\right)-i\left(\mathcal{A}_{\varphi}^{a} \tau_{a}+\mathcal{B}_{\varphi} \tau_{3}\right)\left(\partial_{R} \mathcal{B}_{\mu}+\mathcal{F}_{R \mu}^{(\mathcal{B})}\right) \\
&\left.+i \mathcal{B}_{R}\left(\partial_{\varphi}\mathcal{A}_{\mu}^{b}+\mathcal{F}_{\varphi \mu}^{b}\right) \tau_{b}-i \mathcal{B}_{\varphi}\left(\partial_{R} \mathcal{A}_{\mu}^{b}+\mathcal{F}_{R \mu}^{b}\right) \tau_{b}\right],
\end{aligned}\end{equation}
where $a, b=0,1,2$, and $\eta_{A B}=\operatorname{diag}(-1,1,1,-1)$ and $\tau$s are the generators of $U(1,1)$. Notice that the noncommutative extension produces two extra gauge fields $\mathcal{B}_\mu^{\pm}$, and in order to match the result of the noncommutative version with the commutative limit, we must have the condition 
\begin{equation}\label{0}
\doperator \mathcal{B}_\mu^{(\pm)}=0, 
\end{equation}
 so that the corresponding field strengths to the gauge fields $\mathcal{B}_{\mu}$ vanish. Therefore, in \cite{kawamoto2018charged}, the extra gauge fields $\mathcal{B}_\mu^{\pm}$ were chosen as $\mathcal{B}_{\mu}^{(\pm)}=B \doperator \varphi$ ($B$ is a constant), which simplifies \eqref{2} further, and leads to
\begin{equation}
\begin{aligned}\label{NCgauge}
\mathbb{A}_{\mu}^{(\pm) a} &=-\frac{\theta B}{2}\left[\partial_{R} \mathcal{A}_{\mu}^{(\pm) a}+\mathcal{F}_{R \mu}^{a}\right], \\
\mathbb{B}_{\mu}^{(\pm)} &=-\frac{\theta}{2} \eta_{a b}\left[\mathcal{A}_{R}^{(\pm) a} \mathcal{F}_{\varphi \mu}^{b}-\mathcal{A}_{\varphi}^{(\pm) a} \mathcal{F}_{R \mu}^{b}-\mathcal{A}_{\varphi}^{(\pm) a} \partial_{R} \mathcal{A}_{\mu}^{(\pm) b}\right].
\end{aligned}
\end{equation}
By replacing the vielbeins \eqref{vielbein} and spin connections \eqref{spinc} in \eqref{CSgauge} followed by a noncommutative correction \eqref{NCgauge}, one can obtain the final forms of the CS gauge fields for noncommutative space, $\tilde{A}^{\pm a}$. The metric of the corresponding black hole can be obtained by reconstructing the vielbeins and spin connections as
\begin{equation}
\tilde{e}^a=\frac{\ell}{2}\left(\tilde{A}^{(+)a}-\tilde{A}^{(-)a}\right), \quad \tilde{\omega}^a=\frac{1}{2}\left(\tilde{A}^{(+)a}+\tilde{A}^{(-)a}\right),
\end{equation} 
such that the metric of a charged rotating noncommutative BTZ black hole turns out to be \cite{kawamoto2018charged}
\begin{equation}
\begin{aligned}\label{1}
d s^{2} =&-\left(f(r)-\theta B \frac{2 r^{2}-\epsilon}{4 \ell^{2} r^{2}}\right)\doperator t^{2}+\left(\frac{1}{g(r)}+\theta B \frac{2 f(r) \ell^{2}+2 r^{2}-\epsilon}{4 \ell^{2} r^{2} g^2(r)}\right)\doperator r^{2}\\
&\quad\quad+\left(r^{2}-\frac{\theta B}{2}\right)\doperator \varphi^{2}-\frac{2 \gamma}{\ell} \doperator t \doperator\varphi +\mathcal{O}\left(\theta^{2}\right).
\end{aligned}
\end{equation}

\subsection{Using the fuzziness of mass and charge distributions}\label{sec22}
It has been shown that the space-time noncommutativity eliminates the point-like sources by the smearing of an object over the space \cite{Seiberg_Witten,Dey_Fring_Gouba}, which; in turn, affects the propagation of energy and momentum  of the object \cite{smailagic2002isotropic}. The fuzziness induced by the noncommutativity, thus, clearly results in the modifications of the distributions of mass and charge of the particle. Therefore, an alternative and straightforward way to obtain the metric of a noncommutative BTZ black hole is by replacing the Dirac's point-like mass and charge distributions with a distribution function whose width is the same as the minimal length of the corresponding noncommutative background \cite{nicolini2006noncommutative,Rizzo,Nozari2008,liang2012thermodynamics,Rahaman2013btz}. As for example, the minimal length corresponding to the SW noncommutativity $[x^\mu,x^\nu]=i\theta^{\mu\nu}$ being $\sqrt{\theta}$, one can incorporate such a noncommutativity in mass and charge distributions by considering the corresponding densities to be a Gaussian of width $\sqrt{\theta}$ 
\begin{equation}\label{Densities}
\rho_{c}^{(n+1)D}(r)=\frac{q}{(4\pi\theta)^{n/2}} e^{-r^{2} / 4 \theta}, \quad
\rho_{m}^{(n+1)D}(r)=\frac{M}{(4\pi\theta)^{n/2}} e^{-r^{2} / 4 \theta}.
\end{equation}
Here, $q,M$ and $n$ are the charge, mass and the number of spatial dimensions of the object, respectively. A black hole solution in $AdS_{3}$ space follows from the solution of the Einstein-Maxwell equations (in $8G=c=1$ unit)
\begin{eqnarray}
&& R_{\mu\nu}-\frac{1}{2} g_{\mu\nu} R =\pi\left(T_{\mu\nu}^{\text{matter}}+T_{\mu\nu}^{\text{em}}\right)+\frac{1}{\ell^{2}} g_{\mu \nu}, \label{FEQ}\\
&& \qquad\qquad\quad \frac{1}{\sqrt{-g}} \partial_{\mu}\left(\sqrt{-g} F^{\mu \nu}\right)=J^{\nu}, \label{MaxEQ}
\end{eqnarray}
where the notations are taken to be standard. Since the charge distribution is assumed to be static, the nonzero component of the four current, $J^\nu(r)=\rho_c(r)\delta^\nu_0$, is the charge density only, and the non-vanishing terms in the field strength are $F^{r 0}=-F^{0r}=E(r)$. Now, the continuity equation $T^{\mu\nu}{}_{;\nu}=0$ suggests that the matter distribution be anisotropic so that the most general form of the matter energy-momentum tensor becomes \cite{herrera1997}
\begin{equation}\label{MattTensor}
T_{\mu \nu}^{\text{matter}}=\left(\rho+p_{t}\right) u_{\mu} u_{\nu}+p_{t} g_{\mu \nu}+\left(p_{r}-p_{t}\right) \chi_{\mu} \chi_{\nu},
\end{equation}
where $\rho$ is the energy density and $p_{t}$ and $p_{r}$ are the tangential and radial pressure, respectively. Here, $u^{i}$ and $\chi^{i}$ represent the $(2+1)D$ velocity and the unit vector in the radial direction, respectively. The electromagnetic stress-energy tensor is taken to be of the standard form
\begin{equation}\label{EMTensor}
T_{\mu \nu}^{\text{em}}=-\frac{1}{4 \pi}\left(F_{\mu \alpha} g^{\alpha \beta} F_{\beta \nu}-\frac{1}{4} g_{\mu \nu} F_{\sigma \alpha} g^{\alpha \beta} F_{\beta \rho} g^{\rho \sigma}\right).
\end{equation}
Now, to solve the set of Einstein-Maxwell equations, let us assume a spherically symmetric solution of the form
\begin{equation}\label{MetricAssume}
d s^{2}=-f(r) \doperator t^{2}+[g(r)]^{-1} \doperator r^{2}+r^{2} \doperator \phi^{2},
\end{equation}
with $f(r)=e^{2 \alpha(r)}$ and $[g(r)]^{-1}=e^{2 \beta(r)}$. By using the energy-momentum tensors given by \eqref{MattTensor} and \eqref{EMTensor} as well as the metric in \eqref{MetricAssume}, the field equations given by \eqref{FEQ} can be rewritten as
\begin{eqnarray}\label{4.1}
&& \frac{\beta^{\prime}(r) e^{-2 \beta(r)}}{r} =\pi \rho-\frac{1}{\ell^{2}}+\frac{E^{2}(r)}{8}e^{2[\alpha(r) + \beta(r)]} , \quad \frac{\alpha^{\prime}(r) e^{-2 \beta(r)}}{r}  =\pi p_{r}+\frac{1}{\ell^{2}}-\frac{E^{2}(r)}{8} e^{2[\alpha(r) + \beta(r)]} \notag\\
&& e^{-2 \beta(r)}\left\{[\alpha^{\prime}(r)]^2+\alpha^{\prime \prime}(r)-\alpha^{\prime}(r) \beta^{\prime}(r)\right\}  =\pi p_{t}+\frac{1}{\ell^{2}}+\frac{E^{2}(r)}{8} e^{2[\alpha(r) + \beta(r)]}.
\end{eqnarray}
The equations in \eqref{4.1} suggest that $\alpha(r)=-\beta(r)$ and, therefore, one can use \eqref{MetricAssume} to solve the associated smeared electric field from \eqref{MaxEQ} as
\begin{equation}\label{ElectricF}
E(r)=\frac{1}{r} \int_{0}^{r} \tilde{r} \rho_{c}^{3D}\left(\tilde{r}\right) \doperator \tilde{r}=\frac{q}{2 \pi r}\left(1-e^{-r^{2} / 4 \theta}\right),
\end{equation}
which when is utilized for solving the Einstein-Maxwell equations given by \eqref{4.1} and \eqref{MaxEQ}, we obtain the line element \eqref{MetricAssume} with
\begin{equation}\label{4.12}
e^{-2\beta(r)}=g(r)=f(r)=- M\left(1-e^{-r^{2} / 4 \theta}\right)+\frac{r^{2}}{\ell^{2}}-\frac{q^{2}}{16 \pi^2}\left[\ln |r|+\Gamma \left(0,\frac{r^2}{4\theta}\right)-\frac{1}{2}\Gamma \left(0,\frac{r^2}{2\theta}\right)\right].
\end{equation}
Here, $\Gamma \left(0,x\right)$ denotes an upper incomplete gamma function.
%%%%%%%%%%%%%%%%%%%%%%%%%%%%%%%%%%%%%%%%%%%%%%%%%%%%%%%%%%%%%%%%%%%%%%%%%%%%%%
% Section 3
%%%%%%%%%%%%%%%%%%%%%%%%%%%%%%%%%%%%%%%%%%%%%%%%%%%%%%%%%%%%%%%%%%%%%%%%%%%%%%
\section{Dynamic noncommutative BTZ black holes}\label{sec3}
In this section, we shall use the procedure described in Sec.\,\ref{sec21} to obtain non-static and non-stationary solutions of the BTZ black hole. In Sec.\,\ref{sec21}, the gauge fields $\mathcal{B}_\mu^\pm$ were chosen to be constants for simplicity, so that the corresponding field strengths can vanish and one obtains a simpler solution. However, since the choice of gauge fields $\mathcal{B}_\mu^\pm$ is in our hands, we can explore more intriguing phenomena with the other choices. For instance, we can choose them to be a function of $\varphi$, which yields additional correction terms over \eqref{NCgauge} as follows
\begin{equation}\label{choice1}
\tilde{\mathbb{A}}_{\mu}^{a\pm}= \frac{\theta }{2}\left[(\partial_{\varphi} \mathcal{B}_{\mu})\mathcal{A}_{R}^{a\pm}\right].
\end{equation}
Note that, while making the above-mentioned choice, we have ensured that the commutative limit remains unaltered with the new choice. Clearly, the equation \eqref{choice1} is nonzero only for $a=1$ and, therefore, the extra effective correction terms that can be considered for subsequent discussion takes the form 
\begin{equation}
\tilde{\mathbb{A}}_{\varphi}^{1\pm}= \frac{\theta }{2}\left[(\partial_{\varphi} \mathcal{B}_{\varphi})\mathcal{A}_{R}^{1\pm}\right].
\end{equation}
Obviously, it modifies one of the vielbeins, $e^{1}$, and the exact modified forms of all the vielbeins are as follows
\begin{equation}
\begin{array}{l}
\tilde{e}^{0}=\left(\sqrt{f(r)}-\theta B \frac{2 r^{2}-\epsilon}{8 \ell^{2} r^{2} \sqrt{f(r)}}\right) \doperator t+\frac{\gamma}{\ell \sqrt{f(r)}}\left(1+\theta B \frac{2 r^{2}-\epsilon}{8 f(r) \ell^{2} r^{2}}\right) \doperator \varphi, \\
\tilde{e}^{1}=\left[\frac{1}{\sqrt{g(r)}}+\theta B \frac{2 \ell^{2} f(r)+2 r^{2}-\epsilon}{8 \ell^{2} r^{2} [g(r)]^{3 / 2}}\right] \doperator r +\left[(\partial_{\varphi} {B})\frac{\theta}{2\sqrt{g(r)}}\right]\doperator\varphi, \\
\tilde{e}^{2}=\left[r \sqrt{\frac{g(r)}{f(r)}}-\theta B \frac{2 \ell^{4} r^{2} [f(r)]^{2}-\left(2 r^{2}-\epsilon\right) \gamma^{2}}{8 \ell^{4} r^{3} [f(r)]^{3 / 2} [g(r)]^{1 / 2}}\right] \doperator \varphi.
\end{array}
\end{equation}
Thus, the corresponding metric can be computed as earlier
\begin{eqnarray}
d s^{2}=&& -(\tilde{e}^0)^2+(\tilde{e}^1)^2+(\tilde{e}^2)^2=  -\left[f(r)-\theta B \frac{2 r^{2}-\epsilon}{4 \ell^{2} r^{2}}\right] \doperator t^{2}+\left[r^{2}-\frac{\theta B}{2}\right] \doperator \varphi^{2} \label{10}\\
&& +\left[\frac{1}{g(r)}+\theta B \frac{2 f(r) \ell^{2}+2 r^{2}-\epsilon}{4 \ell^{2} r^{2} g^2(r)}\right] \doperator r^{2}-\frac{2 \gamma}{\ell} \doperator t \doperator \varphi+\left[\theta(\partial_{\varphi} {B})\frac{1}{g(r)}\right] \doperator r \doperator\varphi+\mathcal{O}\left(\theta^{2}\right), \notag
\end{eqnarray}
which is a function of $\varphi$. Now, the metric obtained here is in an unusual and unfamiliar form. The physical implications of such a metric are unclear to us at this moment. However, what we notice is that the metric \eqref{10} is a consequence of a specific dynamic gauge condition. Note that, the usage of the dynamic gauge condition was reported earlier for some models originating from the commutative space-time, see; for instance \cite{alcubierre2001}. Here, we utilize a similar type of gauge condition in the noncommutative space giving rise to metric \eqref{10}. Let us now study another appealing dynamic gauge choice, namely $\mathcal{B}_{t}=C\varphi$, $\mathcal{B}_{r}=0$ and $\mathcal{B}_{\varphi}=Ct$, with $C$ being a constant. The choice itself ensures that the fields do not violate \eqref{0}. In such a case, the vielbeins become
\begin{equation}
\begin{aligned}
\tilde{e}^{0} &=\left(\sqrt{f(r)}-\theta Ct \frac{2 r^{2}-\epsilon}{8 \ell^{2} r^{2} \sqrt{f(r)}}\right) \doperator t+\frac{\gamma}{\ell \sqrt{f(r)}}\left(1+\theta Ct \frac{2 r^{2}-\epsilon}{8 f(r) \ell^{2} r^{2}}\right) \doperator \varphi, \\
\tilde{e}^{1} &=\left[\frac{1}{\sqrt{g(r)}}+\theta Ct \frac{2 \ell^{2} f(r)+2 r^{2}-\epsilon}{8 \ell^{2} r^{2} [g(r)]^{3 / 2}}\right] \doperator r+\left[\frac{\theta C}{2\sqrt{g(r)}}\right]\doperator t, \\
\tilde{e}^{2} &=\left[r \sqrt{\frac{g(r)}{f(r)}}-\theta Ct \frac{2 \ell^{4} r^{2} [f(r)]^{2}-\left(2 r^{2}-\epsilon\right) \gamma^{2}}{8 \ell^{4} r^{3} [f(r)]^{3 / 2} [g(r)]^{1 / 2}}\right] \doperator \varphi,
\end{aligned}
\end{equation}
which render the metric as
\begin{eqnarray}
d s^{2}=&&-\left[f(r)-\theta Ct \frac{2 r^{2}-\epsilon}{4 \ell^{2} r^{2}}\right] \doperator t^{2}+\left[\frac{1}{g(r)}+\theta Ct \frac{2 f(r) \ell^{2}+2 r^{2}-\epsilon}{4 \ell^{2} r^{2} g^2(r)}\right] \doperator r^{2}+\left[r^{2}-\frac{\theta Ct}{2}\right] \doperator \varphi^{2} \notag\\
&&-\frac{2 \gamma}{\ell} \doperator t \doperator \varphi+\left[\frac{\theta C}{g(r)}\right] \doperator r \doperator t+\mathcal{O}\left(\theta^{2}\right).\label{4}
\end{eqnarray}
Here, we can see that the metric becomes non-static as well as non-stationary. Again, a new dynamic gauge condition yields another interesting type of black hole, which is both non-static and non-stationary. It indicates that a better understanding of the extra degree of freedom on the gauge fields is necessary as it can alter the properties of the noncommutative BTZ black hole drastically. The non-static property is a generalization to the rotating black holes where the time-reversal symmetry is broken. Therefore, such a property seemingly does not provide any further significance. However, non-stationary black holes are very interesting objects in the sense that they may provide an inflationary description of the cosmological black holes in which the observable universe is described as the interior of a black hole \cite{pathria}. Since the quantum fluctuation reinforces the concept of noncommutative space-time, a non-stationary noncommutative black hole may resemble an inflationary black hole universe. Moreover, non-stationary metrics are known to be more realistic to describe radiation phenomena in black holes.
%%%%%%%%%%%%%%%%%%%%%%%%%%%%%%%%%%%%%%%%%%%%%%%%%%%%%%%%%%%%%%%%%%%%%%%%%%%%%%
% Section 4
%%%%%%%%%%%%%%%%%%%%%%%%%%%%%%%%%%%%%%%%%%%%%%%%%%%%%%%%%%%%%%%%%%%%%%%%%%%%%%
\section{Charged noncommutative BTZ Black Hole with Lorentzian Distribution}\label{sec4}
In Sec.\,\ref{sec22}, we have discussed a detailed method of obtaining a black hole solution using non-localized distribution functions. While the Gaussian distribution function has been popular in many studies \cite{nicolini2006noncommutative,larranaga2010three,Rahaman2013btz} in this context, there are some enchanting studies arsing from the Lorentzian distribution also \cite{Rizzo,Nozari2008,liang2012thermodynamics}. To our knowledge, most of the studies with the Lorentzian distribution have been carried out for uncharged black holes. The study of Lorentzian charged noncommutative BTZ black hole being inadequate in the literature, we set out to obtain a solution for the same. We shall use the method described in Sec.\,\ref{sec22} and replace the Gaussian distribution functions for the mass and charge densities \eqref{Densities} with the following Lorentzians in $(2+1)D$
\begin{figure}[h]
\centering
\includegraphics[scale=0.8]{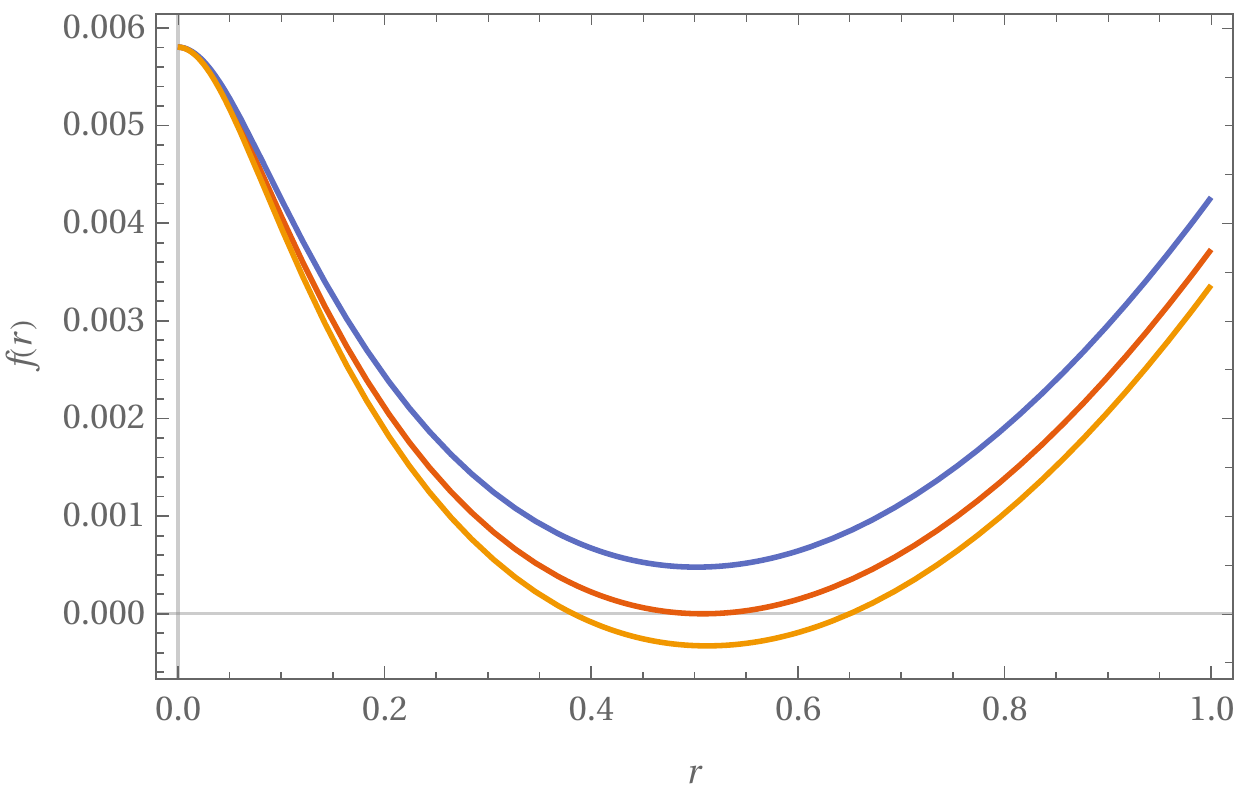}
\caption{\small{Variation of $f(r)$ \eqref{4.11} with $r$ for $q=1,~\ell=10$ and $\theta=0.01$. The purple, red and orange lines correspond to the cases of $ M=0.005000, 0.005593$ and $0.006000$, respectively.}}
\label{fig:4.2}
\end{figure}
\begin{equation}
\rho_{c}^{(2+1)D}(r)=\frac{q \sqrt{\theta}}{2 \pi\left(r^{2}+\theta\right)^{\frac{3}{2}}}, \qquad \rho_{m}^{(2+1)D}(r)=\frac{M \sqrt{\theta}}{2 \pi\left(r^{2}+\theta\right)^{\frac{3}{2}}}.
\end{equation}
The corresponding electric field turns out to be
\begin{equation}
E^{(2+1)D}(r)=\frac{q}{2 \pi r}\left(1-\frac{\sqrt{\theta}}{\sqrt{r^2+\theta}}\right).
\end{equation}
Now, by following the same procedure as described in \eqref{FEQ}-\eqref{4.12}, we obtain a solution of the charged noncommutative BTZ black hole as follows
\begin{equation}\label{4.11}
 e^{-2\beta(r)}=f(r)=-M\left(1-\frac{\sqrt{\theta}}{\sqrt{r^2+\theta}}\right)+\frac{r^{2}}{\ell^{2}}-\frac{q^{2}}{8 \pi^2}\left[\ln (\sqrt{r^{2}+\theta}+\sqrt{\theta})-\frac{\ln \left(\left|r^{2}+\theta\right|\right)}{4}\right].    
\end{equation}
To obtain further information on the derived metric, we study the behavior of the function $f(r)$ with the radius of the black hole. The results are depicted in Fig.\,\ref{fig:4.2}, from where it is obvious that the number of horizons of the black hole depends on the mass of the black hole. If the mass is equal to the extremal black hole, there is exactly one horizon. If the mass exceeds this value, it has two horizons and if it is less, no horizon exists at all.
%%%%%%%%%%%%%%%%%%%%%%%%%%%%%%%%%%%%%%%%%%%%%%%%%%%%%%%%%%%%%%%%%%%%%%%%%%%%%%
% Section 5
%%%%%%%%%%%%%%%%%%%%%%%%%%%%%%%%%%%%%%%%%%%%%%%%%%%%%%%%%%%%%%%%%%%%%%%%%%%%%%
\section{Thermodynamics}\label{sec5}
In this section, we shall analyze some of the thermodynamic properties of various types of noncommutative BTZ black holes that we have derived in earlier sections. Before delving into such an analysis, we introduce the underlying methodology beforehand in the first two subsections.

\subsubsection{Hawking radiation via tunneling formalism}\label{sec511}
By following the approaches described in \cite{parikh2000hawking,arzano2005hawking, banerjee2008quantum,Nozari2008,chougule2018btz,anacleto2018quantum}, we shall use the tunneling formalism to study the thermodynamics of the black holes under consideration. Among various other methods, it is a relatively new and convenient method as outlined, for instance, in \cite{chougule2018btz}. On many occasions, for a spherically symmetric metric, there exists a coordinate singularity at its outer horizon $r_+$. Therefore, the study of physics at the event horizon requires a transformation of the coordinate system beforehand, so that the metric becomes well-behaved at the horizon. Thanks to the Painlev\'e transformation 
\begin{equation}
\doperator t \rightarrow \doperator t'-\sqrt{\frac{1-g(r)}{f(r) g(r)}} \doperator r,
\end{equation}
which essentially removes the coordinate singularity and regularizes the black hole at its horizon. Now, the null geodesics for the modified regularized metric
\begin{equation}
d s^{2}=-f(r) \doperator t'^{2}+2 f(r) \sqrt{\frac{1-g(r)}{f(r) g(r)}} \doperator t' \doperator r+\doperator r^{2}+r^{2} d \Omega^{2},
\end{equation}
takes the form
\begin{equation}
\dot{r} \equiv \deriv{r}{t'}=\sqrt{\frac{f(r)}{g(r)}}\left(\pm 1-\sqrt{1-g(r)}\right).
\end{equation}
Since we are interested in tunneling through the horizon only, we can use the Taylor expansion of the radial null geodesics for $f$ and $g$ in the near horizon region
\begin{equation}
\dot{r}=\frac{1}{2} \sqrt{f^{\prime}\left(r_+\right) g^{\prime}\left(r_+\right)}\left(r-r_+\right)+\mathcal{O}\left(\left(r-r_+\right)^{2}\right),
\end{equation}
and the surface gravity of the metric can be computed as
\begin{equation}
\mathcal{K}(M)=\left.\Gamma_{00}^{0}\right|_{r=r_+}=\left.\frac{1}{2}\left(\sqrt{\frac{g(r)-g^2(r)}{f(r)}} \deriv{f(r)}{r}\right)\right|_{r=r_+} \simeq \frac{1}{2} \sqrt{f^{\prime}\left(r_+\right) g^{\prime}\left(r_+\right)}.
\end{equation}
The Hawking temperature in the tunneling formalism is then given by
\begin{equation}\label{5.1}
T_H=\frac{\mathcal{K}}{2 \pi}=\frac{1}{4 \pi} \sqrt{f^{\prime}\left(r_+\right) g^{\prime}\left(r_+\right)},
\end{equation}
which can also be written in the form of surface gravity.

\subsubsection{Entropy}\label{sec52}
The entropy of a charged rotating black hole can be derived from the first law of thermodynamics
\begin{equation}\label{1stTherm}
\doperator M=T_H \doperator S+\Omega \doperator J+\Phi\doperator q.
\end{equation}
Since, the total mass $M$, angular velocity $\Omega$ and the electric potential $\Phi$ at the outer horizon are defined as $M=M(r_+,J,q)$, $\Omega=\partial M/\partial J$ and $\Phi=\partial M/\partial q$, respectively, we obtain
\begin{equation}
\doperator M=\pderiv{M}{r_+} \doperator r_++\Omega \doperator J+\Phi\doperator q,
\end{equation}
which when compared with \eqref{1stTherm}, the entropy turns out to be
\begin{equation}\label{2.2}
\doperator S=\frac{1}{T_{H}}\pderiv{M}{r_+}\doperator r_+.
\end{equation}
The total mass $M$ of the black hole can be calculated from the metric. For example, for a metric of the form \eqref{MetricAssume}, the mass can be computed from the condition that $g(r)$ vanishes at the outer horizon. Thus, the knowledge of the total mass $M$ and the Hawking temperature $T_H$ will facilitate us to obtain an exact expression of the entropy. 

\subsection{Thermodynamics of charged noncommutative CS BTZ black hole}
In this section, we shall study the thermodynamic properties of the metric given by \eqref{1}, which corresponds to a charged rotating noncommutative BTZ black hole. Although, the same for the uncharged case has been studied earlier in \cite{anacleto2018quantum}, however; to our knowledge, the thermodynamics of the charged case has not been explored yet. Recall that the Hawking temperature in the tunneling formalism, as described in Sec.\,\ref{sec511}, applies to the metric having a spherically symmetric form only. However, our metric \eqref{1} is not in the required form and, thus, it claims a coordinate transformation
\begin{figure}[h]
\subfigure[]{
\includegraphics[scale=0.64]{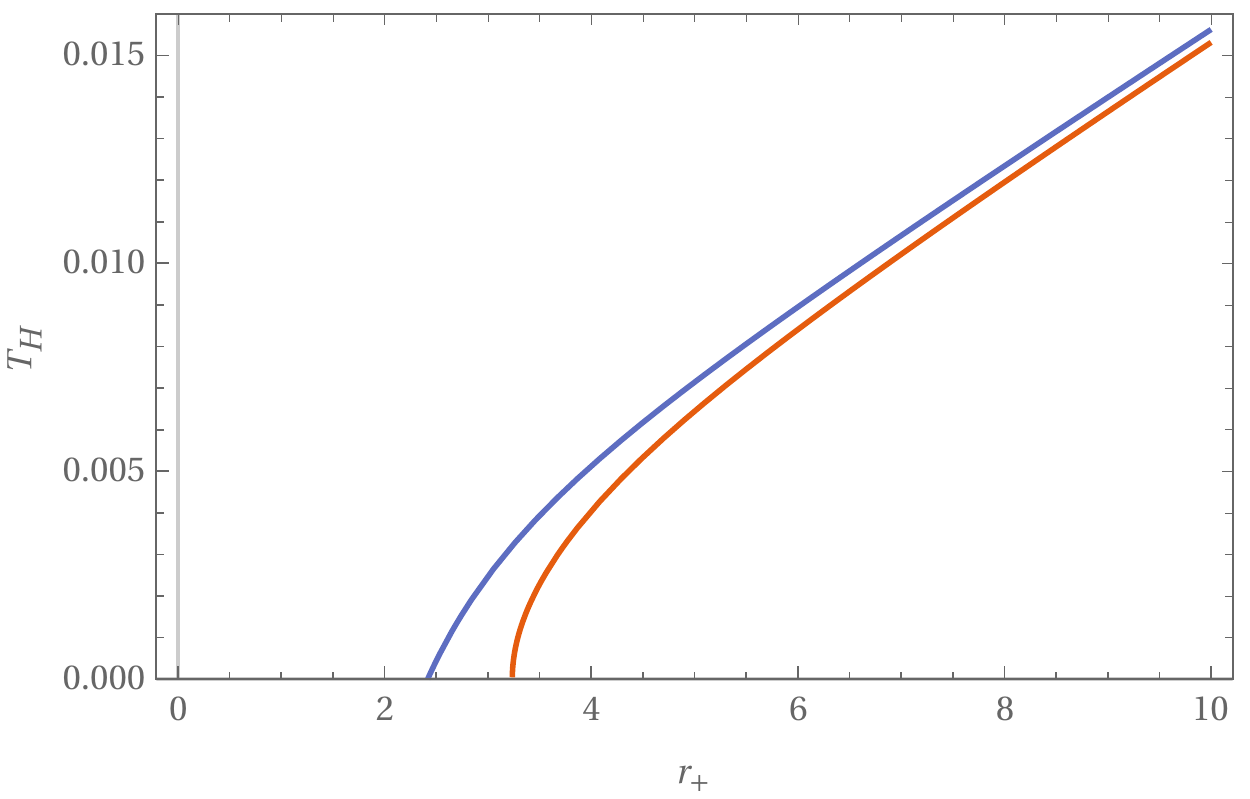}
\label{fig:4.4a}}
\subfigure[]{
\includegraphics[scale=0.64]{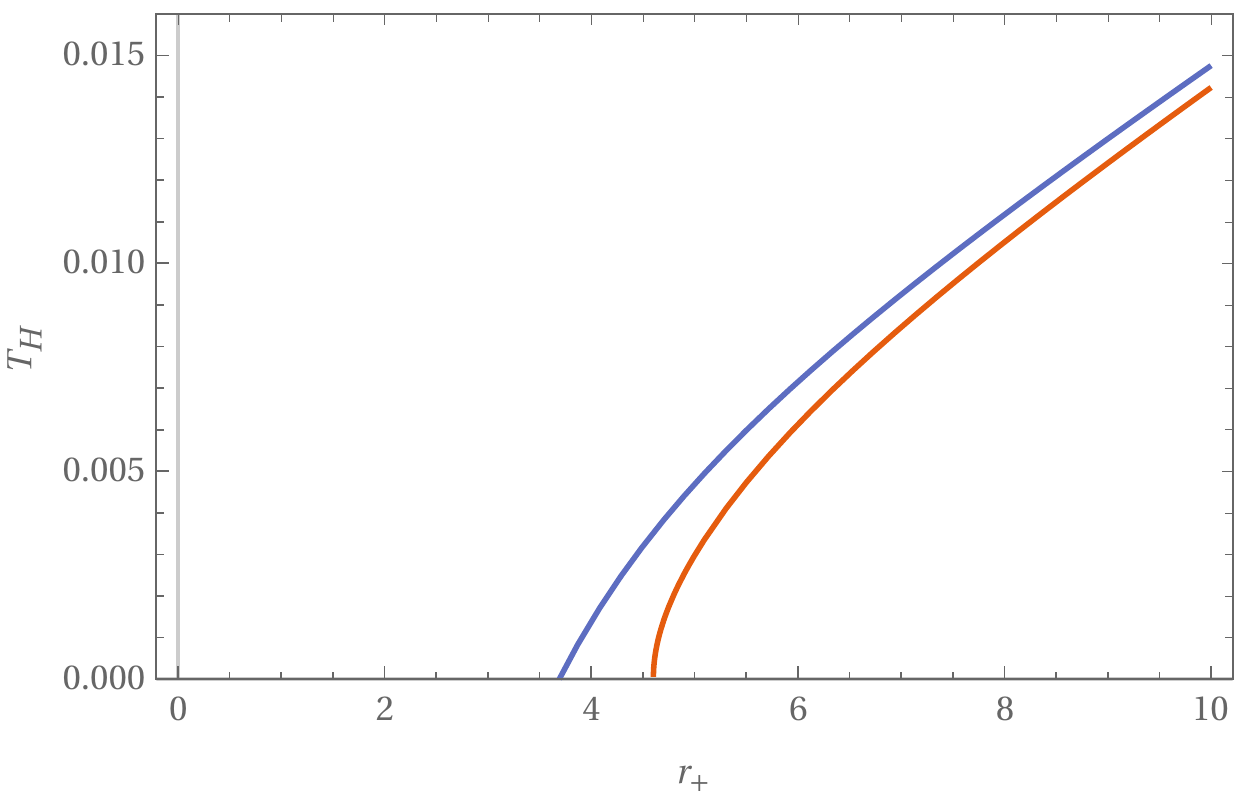}
\label{fig:4.4b}}
\caption{\small{$T_H$ as a function of $r_{+}$ for $\ell=10$ \subref{fig:4.4a} $\theta B=8,~q=0.1$ and $J=1$ \subref{fig:4.4b} $\theta B=15,~q=0.2$ and $J=2$ The orange and purple lines represent the noncommutative and commutative cases, respectively.}}
\label{fig:4.4}
\end{figure}
\begin{equation}\label{j}
\doperator \varphi\rightarrow\doperator \varphi'+\frac{J}{2 r^2-\theta B}\doperator t,
\end{equation}
so that the transformed spherically symmetric metric in natural units ($8G=c=1$) takes the form 
\begin{equation}\label{BH10}
d s^{2}=-\left(\mathcal{G}(r)-M+\frac{\theta BJ^2}{8r^4}\right)\doperator t^{2}+\left(\mathcal{G}(r)-M+\frac{\theta B M}{2r^2}-\frac{\theta B}{2\ell ^2}\right)^{-1}\doperator r^{2}+\left(1-\frac{\theta B}{2 r^{2}}\right) r^{2}\doperator \varphi'^{2},
\end{equation}
where
\begin{equation}
\mathcal{G}(r)=\frac{r^{2}}{\ell ^{2}}+\frac{J^{2}}{4 r^{2}}-\mathcal{\pi} q^{2} \ln r -\frac{\theta B}{2 \ell^{2}}+\frac{\theta B \pi q^{2}}{4 r^{2}}.
\end{equation}
Therefore, by following \eqref{5.1}, the Hawking temperature acquires the form 
\begin{equation}\label{3.3}
T_H = \frac{\lambda(r_+)}{2 \pi \ell^{2}} \left[r_+^2-\theta B\lambda^{-1}(r_+)\left(1-\lambda(r_+)+\frac{\ell^2M}{2r_+^2}\right)+\mathcal{O}\left(\theta^{2}\right)\right]^{1/2},
\end{equation}
with 
\begin{equation}
\lambda(r_+)=1-\frac{\ell^{2} J^{2}}{4 r_+^{4}}-\frac{\pi q^{2} \ell^{2}}{2 r_{+}^{2}}.
\end{equation}
Thus, the total mass of the black hole \eqref{BH10} is computed from the condition that at the outer horizon $r_+$, the quantity, $\mathcal{G}(r)-M+\frac{\theta B M}{2r^2}-\frac{\theta B}{2\ell^2}$, vanishes, which implies that
\begin{equation}\label{Mass}
M =\left(1-\frac{\theta B}{2 r_{+}^{2}}\right)^{-1}\left(\mathcal{G}(r_+)-\frac{\theta B}{2\ell^2}\right)=\mathcal{G}(r_+)+\frac{\theta B}{2}\left(\frac{\mathcal{G}(r_+)}{r_+^2}-\frac{1}{\ell^2}\right)+\mathcal{O}\left(\theta^{2}\right).
\end{equation}
Now, by using \eqref{3.3} and \eqref{Mass}, it is straightforward to obtain the entropy by using \eqref{2.2}. However, the expression of the entropy being slightly lengthy, we do not present it explicitly. More importantly, 
our analysis in the present scenario does not require the explicit form of entropy. Therefore, we avoid such an expression, which, anyway does not carry significant importance in our context. A more important entity is the Hawking temperature provided by \eqref{3.3}. The analysis of temperature is shown in Fig.\,\ref{fig:4.4} for two sets of values of the charge, angular momentum and the noncommutative parameter. The orange lines correspond to the noncommutative cases associated with the fixed value of the noncommutative parameter, whereas the purple lines show the same in the commutative limit $\theta\rightarrow 0$. As it turns out that the behavior of the black hole temperature in the noncommutative case is similar to the commutative one, we only obtain a correction of the temperature due to the noncommutativity. We also do not notice any restriction on the noncommutative parameter in the sense that the black hole is physical and behaves normally for all values of the noncommutative parameter.

Thermodynamics of \eqref{4} can be studied in the tunneling formalism but, now in this case, the temperature will be time-dependent. The line element \eqref{4} is not yet in a spherically symmetric form and, therefore, in order to apply the tunneling formalism, we apply the following coordinate transformation
\begin{equation}
\doperator \varphi\rightarrow\doperator \varphi'+\frac{J}{2 r^2-\theta B}\doperator t, \quad \doperator{r}\rightarrow\doperator{r}'-\frac{\theta C}{2} \doperator{t}.
\end{equation}
The line element up to the first-order correction in $\theta$ is, thus, modified as
\begin{equation}
d s^{2}=-\mathcal{F}(r') \doperator t^{2}+\mathcal{H}^{-1}(r') \doperator r'^{2}+\left[r'^{2}-\frac{\theta Ct}{2}\left(1+2r'\right)\right] \doperator \varphi'^{2},
\end{equation}
where
\begin{eqnarray}
&& \mathcal{F}(r')=\frac{r'^{2}}{\ell^{2}}-M+\frac{J^{2}}{4 r'^{2}}-\pi q^{2} \ln r'-\frac{\theta C t\lambda(r'_+)}{2\ell^2}(1+2r')\\
&& \mathcal{H}(r')=\frac{r'^{2}}{\ell^{2}}-M+\frac{J^{2}}{4 r'^{2}}-\pi q^{2} \ln r'-\frac{\theta C t}{2}\left[\frac{2(1+r')}{\ell^2}-\frac{M}{r'^2}-\frac{\pi q^2 (1+2r')}{2r'^2}-\frac{J^2}{2r'^3}\right].
\end{eqnarray}
Since the radius of the horizon $r'$ is obtained as a solution of $\mathcal{H}(r')=0$, it is worthwhile to mention that, interestingly, it becomes time-dependent. Clearly, the Hawking temperature also turns out to be time-dependent
\begin{equation}\label{TD}
T_H = \frac{\lambda(r'_+)}{2 \pi} \left[\frac{r_+^{'2}}{\ell^4}-\frac{\theta Ct\lambda^{-1}(r'_+)}{2\ell^2}\left\{\frac{M}{r_+^{'2}}+\frac{3J^2}{2r_+^{'3}}+\frac{\pi q^2}{r'_+}+\frac{2-2\lambda(r'_+)+2r'_+}{\ell^2}\right\}+\mathcal{O}\left(\theta^{2}\right]\right]^{1/2}.
\end{equation}
It should be noted that the time-dependence of temperature and horizon radius is not a surprising result, because the corresponding metric is non-static and non-stationary. These time-dependent thermodynamic quantities may give rise to dynamic characteristics of the Hawking radiation and evaporation of the black hole. To realize this, we can simplify \eqref{TD} for the chargeless and nonrotating scenario by taking the limits $J\rightarrow 0$ and $q\rightarrow 0$. Thereafter, if we choose $\ell=10,\theta C=1$ and the Hawking temperature $T_H$ to be zero, the equation \eqref{TD} will be left with two free parameters, $r_+$ and $t$. Therefore, we can easily obtain a set of possible values of the horizon radius after the evaporation is complete and the corresponding time taken to evaporate, which are, however, not possible to determine exactly unless we know how the radius of the horizon changes with time. Explicit expressions of time-dependent horizon radius and total mass of a dynamic noncommutative black hole are yet unexplored and out of our scope in the present study. This problem itself is a new and separate project that may require a longer time to explore. One of the possible ways to address the time-dependent mass is to follow the method prescribed in \cite{Page_2013}; however, such a study in a noncommutative background may demand more sophistication and, thus, we keep it as a separate problem to be addressed in future.  

\subsection{Thermodynamics of charged noncommutative Lorentzian BTZ black hole}\label{sec53}
The aim of this section is to study the thermodynamic properties of the metric obtained in \eqref{4.11} using the Lorentzian distribution function. However, before that, in order to introduce an interesting discussion, let us perform the same analysis for the metric \eqref{4.12} associated with the Gaussian distribution function. The semi-classical Hawking temperature as per \eqref{5.1} turns out to be
\begin{figure}[h]
\subfigure[]{
\includegraphics[scale=0.64]{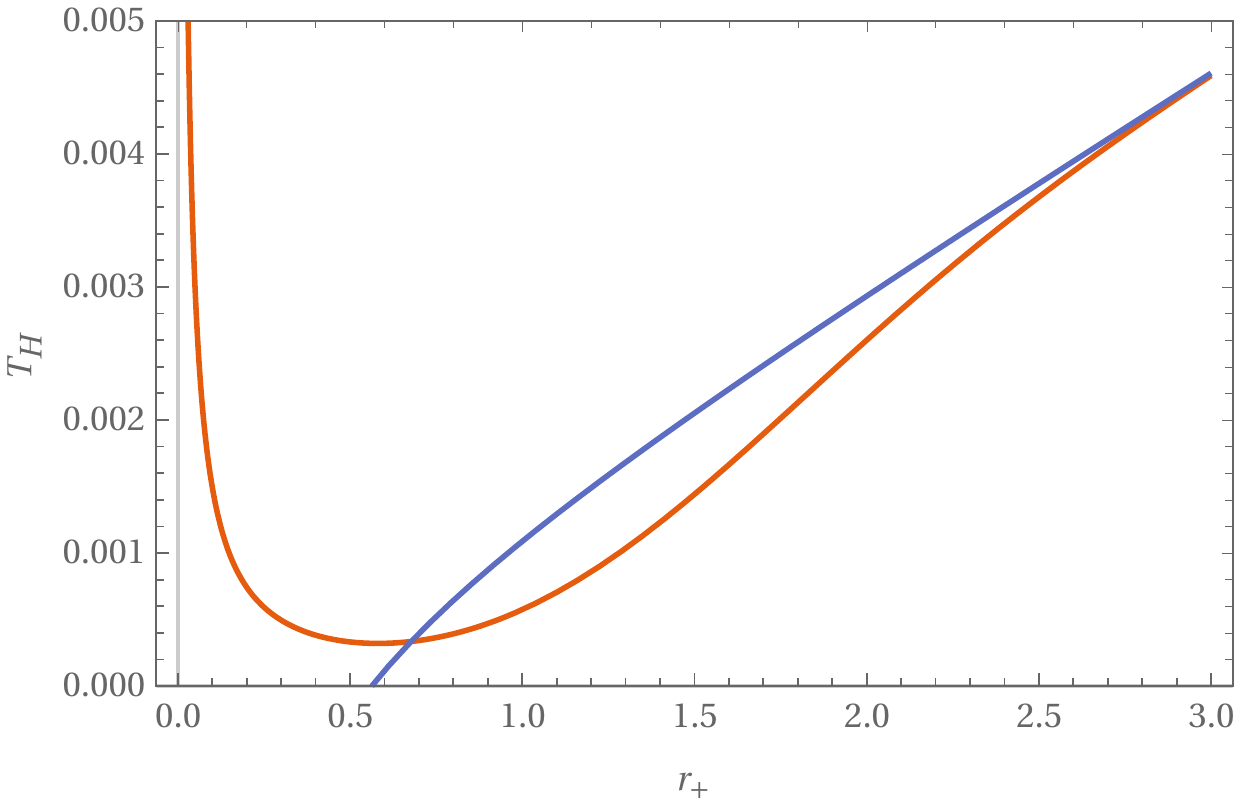}
\label{fig:4.1a}}
\subfigure[]{
\includegraphics[scale=0.64]{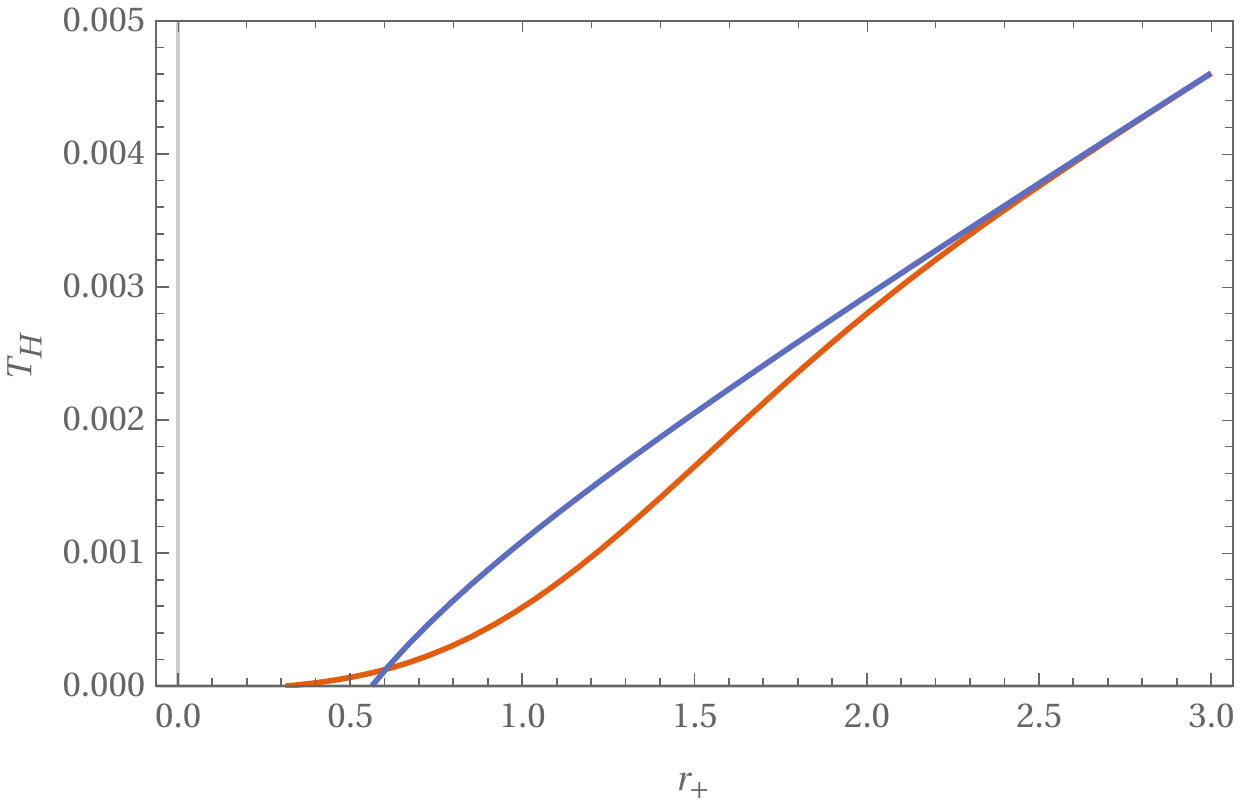}
\label{fig:4.1b}}
\caption{\small{$T_H$ vs $r_+$ for the metric \eqref{4.12} for $\ell=10,~q=1$ \subref{fig:4.1a} $\theta=0.3$ \subref{fig:4.1b} $\theta=0.22$.  The orange and purple lines represent the noncommutative and commutative cases, respectively.}}
\label{fig:4.1}
\end{figure}
\begin{equation}\label{5.8}
T_{H}=\frac{r_{+}}{2 \pi \ell^{2}}\left[1-\frac{ M \ell^{2}}{4\theta} e^{-r_{+}^{2} / 4 \theta}-\frac{q^{2} \ell^{2}}{32 \pi^2 r_{+}^{2}}\left(1+e^{-r_{+}^{2} / 2 \theta}-2 e^{-r_{+}^{2} / 4 \theta}\right)\right],
\end{equation}
whereas the total mass acquires the form
\begin{equation}\label{5.9}
M=\frac{1}{\left(1-e^{-r_{+}^{2} / 4 \theta}\right)}\left[\frac{r_+^{2}}{\ell^{2}}-\frac{q^{2}}{16 \pi^2}\left[\ln |r_+|+\Gamma \left(0,\frac{r_+^2}{4\theta}\right)-\frac{1}{2}\Gamma \left(0,\frac{r_+^2}{2\theta}\right)\right]\right].
\end{equation}
Therefore, the entropy can be calculated easily from \eqref{2.2} as follows
\begin{equation}
S=4\pi \int_{r_{0}}^{r_{+}}\frac{1}{\left(1-e^{-\psi^{2} / 4 \theta}\right)} \doperator \psi.
\end{equation}
The Hawking temperature for the noncommutative charged BTZ black hole with Lorentzian distribution given by \eqref{4.11} is computed from \eqref{5.1} as
\begin{equation}\label{5.6}
T_{H}=\frac{r_{+}}{4 \pi}\left[\frac{2}{\ell^{2}}-\frac{M \sqrt{\theta}}{(r_{+}^{2}+ \theta)^{\frac{3}{2}}}-\frac{q^{2}}{16 \pi^2 r_{+}^2}\left(1-\frac{\sqrt{\theta}}{\sqrt{r_{+}^{2}+\theta}}\right)^{2}\right].
\end{equation}
The total mass of the same being
\begin{equation}\label{5.7}
M=\left(1-\frac{\sqrt{\theta}}{\sqrt{r_{+}^{2}+\theta}}\right)^{-1}\left[\frac{r_{+}^{2}}{\ell^{2}}-\frac{q^{2}}{8 \pi^2}\left\{\ln (\sqrt{r_{+}^{2}+\theta}+\sqrt{\theta})-\frac{\ln \left(\left|r_{+}^{2}+\theta\right|\right)}{4}\right\}\right],
\end{equation}
the entropy from \eqref{2.2} becomes
\begin{equation}
S=4\pi \int_{r_{0}}^{r_{+}}\frac{\sqrt{\psi^2+\theta}}{\sqrt{\psi^2+\theta}-\sqrt{\theta}} \doperator \psi.
\end{equation}
\begin{figure}[h]
\subfigure[]{
\includegraphics[scale=0.64]{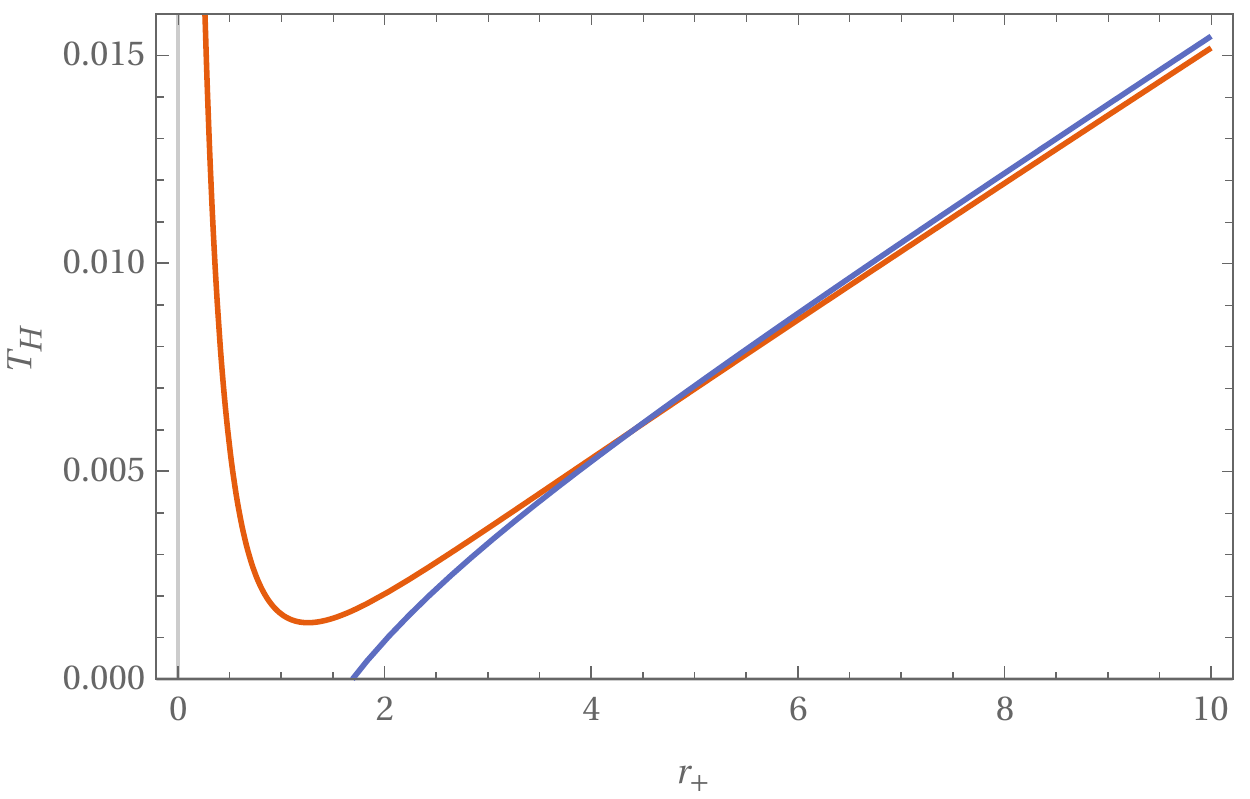}
\label{fig:4.3a}}
\subfigure[]{
\includegraphics[scale=0.64]{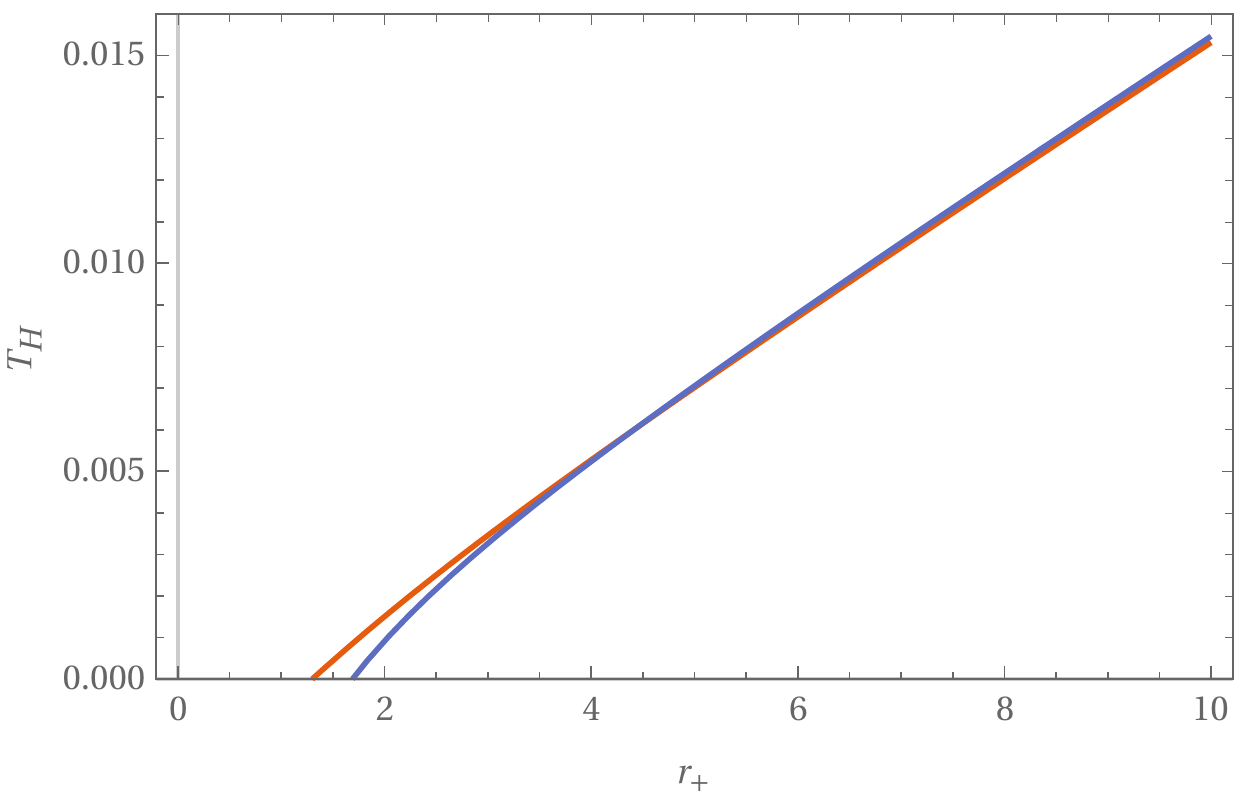}
\label{fig:4.3b}}
\caption{\small{$T_H$ vs $r_+$ for the metric \eqref{4.11} for $\ell=10,~q=3$ \subref{fig:4.3a} $\theta=0.2$ \subref{fig:4.3b} $\theta=0.06$. The orange and purple lines represent the noncommutative and commutative cases, respectively.}}
\label{fig:4.3}
\end{figure}
\begin{figure}[h]
\subfigure[]{
\includegraphics[scale=0.64]{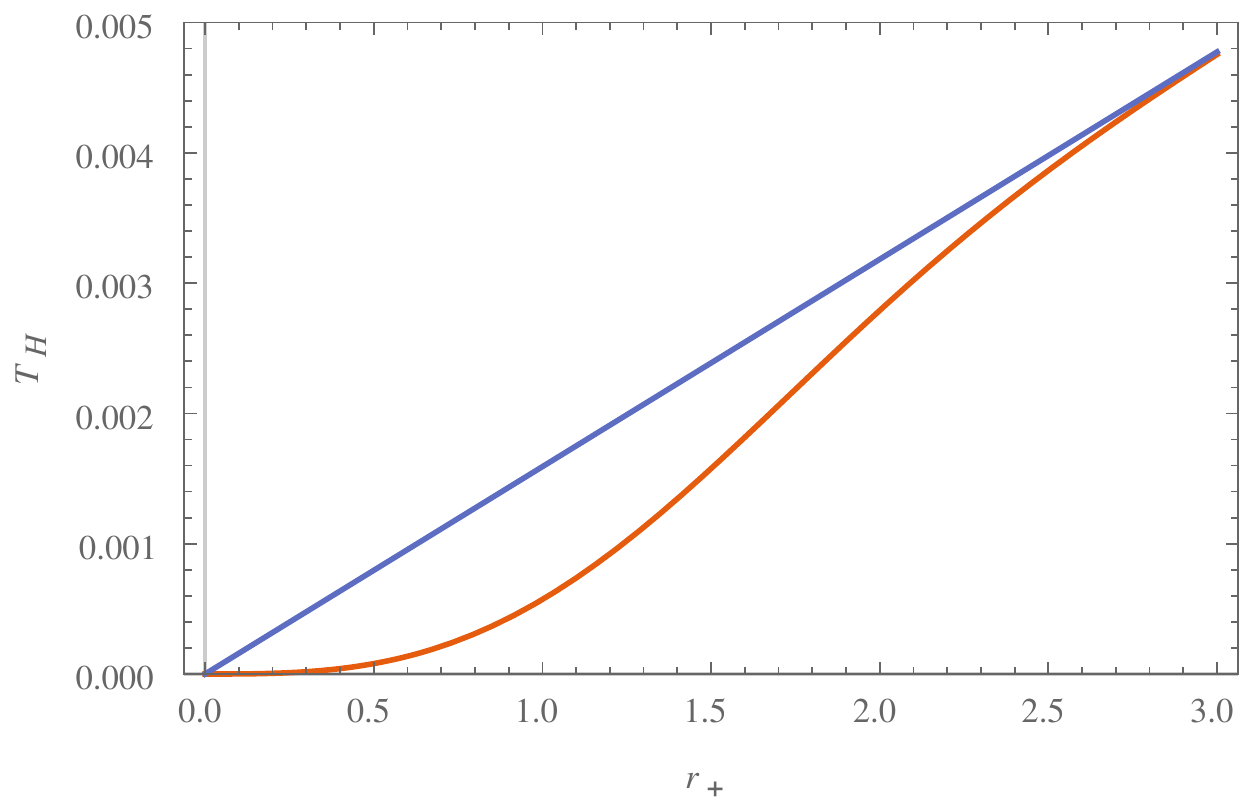}
\label{fig6a}}
\subfigure[]{
\includegraphics[scale=0.64]{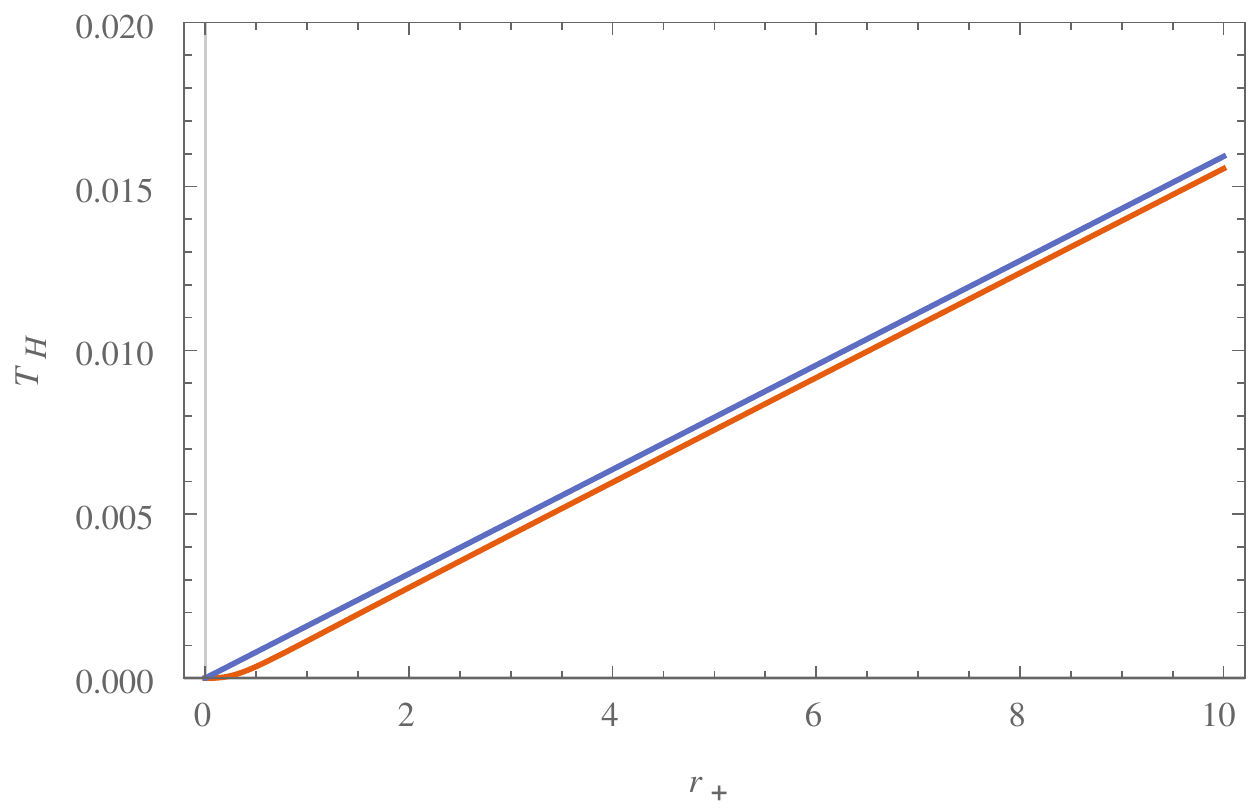}
\label{fig6b}}
\caption{\small{$T_H$ vs $r_+$ for $\ell=10,~q=0$ \subref{fig6a} for the metric \eqref{4.12} with $\theta=0.3$ \subref{fig6b} for the metric \eqref{4.11} with $\theta=0.2$. The orange and purple lines represent the noncommutative and commutative cases, respectively.}}
\label{fig6}
\end{figure} 
The variation of Hawking temperature $T_H$ with the outer horizon radius $r_+$ for the Gaussian and Lorentzian cases are shown in Fig.\,\ref{fig:4.1} and Fig.\,\ref{fig:4.3}, respectively. In all such cases, it is clear that if the radius of the outer horizon decreases, the temperature corresponding to the purple lines (commutative cases) will decrease correspondingly. However, in the case of the orange lines (noncommutative cases), the temperatures show an inverse behavior when the values of $r_+$ are small. Clearly, such a converse and, in fact, the divergent behavior of the temperature is unexpected and nonphysical. However, we notice that the divergent behavior of the temperature can be ceased, provided that the noncommutative parameter is bound within a small region depending on the amount of charge contained by the black hole. For example, in Fig.\,\ref{fig:4.3a}, the value of the noncommutative parameter is $\theta=0.2$ leading to a divergent behavior. Whereas, if we set $\theta=0.06$, as shown in Fig.\,\ref{fig:4.3b}, we can prevent the nonphysical behavior. We find that, in the given case, the critical value of $\theta$ is $0.09$, and the divergence occurs for $\theta>0.09$ (considering a second-order accuracy). For a different value of charge $q$, the bound on $\theta$ will change correspondingly. A similar type of analysis is applicable to the Gaussian case also, which has been depicted in Fig.\,\ref{fig:4.1}, where the noncommutative parameter is bound by $\theta<0.23$. Nevertheless, as it is well-known that the noncommutative parameter is supposed to be bound by small values, and our analysis does not violate the general findings. Rather, we obtain a better estimation of the parameter for which the corresponding black hole will be stable and physical.

Note that, the bound on the noncommutative parameter $\theta$ arises as a distinctive feature of the charged case only. In the chargeless scenario, irrespective of whether it is rotating or nonrotating, one does not find such a bound. For example, the chargeless nonrotating Gaussian and Lorentzian cases can be studied by choosing $q=0$ in \eqref{5.8} and \eqref{5.6}, respectively. The Hawking temperature for these cases are plotted in Fig.\,\ref{fig6a} and \ref{fig6b} in the respective order, and we do not notice any divergent behavior in the temperature and, thus, no bound. The rotating case of a chargeless noncommutative black hole using the Lorentzian distribution has been explored in \cite{liang2012thermodynamics}, which is also in the favor of nonexistence of the bound. Furthermore, We have verified that the chargeless rotating noncommutative black holes emerging from the Gaussian distribution function do not yield the bound. Even though our study of Hawking temperature relies on a semiclassical approach, the short-distance behavior within this approach may be trusted because we are including correction terms due to the noncommutative space-time, which are known to be reliable in the length scale around the Planck scale. A similar prescription was utilized in the study of short-distance features of Hawking temperature for a Schwarzschild black hole in a noncommutative space-time \cite{nicolini2006noncommutative}. Thus, it gives us the confidence that the divergent behavior of the temperature in Fig.\,\ref{fig:4.1} and \ref{fig:4.3} may be purely due to the inclusion of the charge.
%%%%%%%%%%%%%%%%%%%%%%%%%%%%%%%%%%%%%%%%%%%%%%%%%%%%%%%%%%%%%%%%%%%%%%%%%%%%%%
% Section 6
%%%%%%%%%%%%%%%%%%%%%%%%%%%%%%%%%%%%%%%%%%%%%%%%%%%%%%%%%%%%%%%%%%%%%%%%%%%%%%
\section{Conclusions}\label{sec6}
We have studied two different types of charged BTZ black holes in noncommutative spaces emerging from two independent approaches. In the first approach, we incorporate the noncommutativity on the CS gauge theory by using the famous SW map. Whereas, in the second approach, we utilize the fuzziness property of the noncommutativity and introduce the same with a Lorentzian distribution having the same width as the corresponding noncommutativity to be imposed. The novelty of our results is, at least, two-fold. First, we explore the dynamic charged noncommutative BTZ  black holes for the first time in the literature. This is obtained by imposing dynamic gauge freedom to the CS gauge theory, which was mostly ignored earlier. We argue that such a unique choice of gauge may give rise to exotic cases, like; non-static and non-stationary black holes. Such dynamic black holes may cultivate striking features in inflationary black hole cosmological scenarios. Second, we construct the fuzziness-induced noncommutative BTZ black hole solution for the charged case and analyze some of the thermodynamic properties using the tunneling formalism. One of the important notions of the inclusion of the charge to the metric is that we are able to provide a bound on the noncommutative parameter, which is not possible in the first approach. We also notice that such a fact is a consequence of the charged black hole only, and the bound is entirely dependent on the amount of charge the black hole contains. As soon as the charge is removed, it is no longer possible to detect the bound. Therefore, we believe that the construction of the charged BTZ black hole in the noncommutative background brings in further remarkable possibilities.

Clearly, our results add outstanding aspects to the noncommutative extension of the charged BTZ black hole and we believe that further investigation on the dynamic black holes may captivate more unusual characteristics. Since a proper methodology for analyzing the thermodynamic properties of the dynamic black holes of type \eqref{10} is inadequate in the literature, we have not been able to provide the same. However, it will be worth building a suitable mechanism for analyzing the thermodynamic properties of dynamic noncommutative black holes. Furthermore, we have explored the black holes arising from the SW noncommutativity, however, there exist many other types of noncommutativities having different implications in different scenarios and, we believe that the corresponding black holes may provide further insights, especially for the dynamic choices of the gauge. 

\vspace{0.5cm} \noindent \textbf{\large{Acknowledgments:}} S.\,D.\,acknowledges the support of the research grant (DST/INSPIRE/04/2016/ 001391) by DST-INSPIRE, Govt. of India.
%%%%%%%%%%%%%%%%%%%%%%%%%%%%%%%%%%%%%%%%%%%%%%%%%%%%%%%%%%%%%%%%%%%%%%%%%%%%%%
%  References
%%%%%%%%%%%%%%%%%%%%%%%%%%%%%%%%%%%%%%%%%%%%%%%%%%%%%%%%%%%%%%%%%%%%%%%%%%%%%%
%\bibliographystyle{unsrt} 
%\bibliography{Ref.bib}
%\input{Reference.tex}
%-----------------------------------------------------------------------------

%%%%%%%%%%%%%%%%%%%%%%%%%%%%%%%%%%%%%%%%%%%%%%%%%%%%%%%%%%%%%%%%%%%%%%%%%%%%%%
%  Rulling
%%%%%%%%%%%%%%%%%%%%%%%%%%%%%%%%%%%%%%%%%%%%%%%%%%%%%%%%%%%%%%%%%%%%%%%%%%%%%%
%\vspace{0.3cm}
%\begin{center}
%\rule{2.5cm}{1.0pt}
%\end{center}
%%%%%%%%%%%%%%%%%%%%%%%%%%%%%%%%%%%%%%%%%%%%%%%%%%%%%%%%%%%%%%%%%%%%%%%%%%%%%%
\end{document}